\documentstyle[amssymb,preprint,aps]{revtex}

\begin{document}
\draft
\title{Andreev bound states and $\pi $-junction transition in a superconductor /
quantum-dot / superconductor system}
\author{{\ Yu Zhu, Qing-feng Sun and }Tsung-han Lin$^{*}$}
\address{{\it State Key Laboratory for Mesoscopic Physics and }\\
{\it Department of Physics, Peking University,}{\small \ }{\it Beijing}\\
100871, China}
\date{}
\maketitle

\begin{abstract}
We study Andreev bound states and $\pi $-junction transition in a
superconductor / quantum-dot / superconductor (S-QD-S) system by Green
function method. We derive an equation to describe the Andreev bound states
in S-QD-S system, and provide a unified understanding of the $\pi $-junction
transition caused by three different mechanisms: (1) {\it Zeeman splitting.}
For QD with two spin levels $E_{\uparrow }$ and $E_{\downarrow }$, we find
that the surface of the Josephson current $I(\phi =\frac \pi 2)$ vs the
configuration of $(E_{\uparrow },E_{\downarrow })$ exhibits interesting
profile: a sharp peak around $E_{\uparrow }=E_{\downarrow }=0$; a positive
ridge in the region of $E_{\uparrow }\cdot E_{\downarrow }>0$; and a {\em %
negative}, flat, shallow plain in the region of $E_{\uparrow }\cdot
E_{\downarrow }<0$. (2){\it \ Intra-dot interaction.} We deal with the
intra-dot Coulomb interaction by Hartree-Fock approximation, and find that
the system behaves as a $\pi $-junction when QD becomes a magnetic dot due
to the interaction. The conditions for $\pi $-junction transition are also
discussed. (3) {\it \ Non-equilibrium distribution.} We replace the Fermi
distribution $f(\omega )$ by a non-equilibrium one $\frac 12\left[ f(\omega
-V_c)+f(\omega +V_c)\right] $, and allow Zeeman splitting in QD where $%
E_{\uparrow }=-E_{\downarrow }=h.$ The curves of $I(\phi =\frac \pi 2)$ vs $%
V_c$ show the novel effect of interplay of non-equilibrium distribution with
magnetization in QD.
\end{abstract}



\baselineskip 20pt 

\section{INTRODUCTION}

Superconductivity has the nature of quantum condensation in a macroscopic
scale, which can be described by a wavefunction with phase factor $e^{\text{i%
}\phi }$. When two superconductors are weak linked, the phase difference
will manifest in the dc Josephson current, with the current-phase relation $%
I=I_c\sin (\phi _1-\phi _2)$. If the weak link area is controlled by certain
external conditions, the magnitude of the critical current $I_c$ may be
either suppressed \cite{sup} or enhanced \cite{enh}. In some occasions, even
the sign of $I_c$ may be reversed \cite{rev}, or equivalently the phase
factor $\sin (\phi _1-\phi _2)$ changes to $\sin (\phi _1-\phi _2+\pi )$,
referred to as the $\pi $-junction transition. One of the simplest example
is the so-called superconducting quantum diffraction, in which a
superconductor / insulator / superconductor tunnel junction is tuned by an
external magnetic field. The dc Josephson current vs the magnetic flux $\Phi
_B$ has the form $I=I_c\left[ \frac{\sin \pi \Phi _B/\Phi _0}{\pi \Phi
_B/\Phi _0}\right] \sin (\phi _1-\phi _2)$, and changes its sign at every $%
\Phi _B/\Phi _0$ equal to an integer. Unfortunately, the $\pi $-junction
transition cannot be directly detected in the two terminal tunnel junction
because the current source is used in the measurement to control
supercurrent rather than phase difference. However, in a mesoscopic
superconductor / normal-metal / superconductor (SNS) junction with the
N-region coupled to normal electrode(s) \cite{arm1,arm2}, the phase
difference can be determined independently by the coherent Andreev
reflection current which is proportional to $\left| e^{\text{i}\phi _1}+e^{%
\text{i}\phi _2}\right| ^2=2\left[ 1+\cos (\phi _1-\phi _2)\right] $. Thus,
direct observation of the $\pi $-junction transition becomes accessible.

In mesoscopic SNS junctions, supercurrent is conducted through the N-region
by Andreev reflection (AR) process \cite{btk}. The energy gaps of two
superconducting electrodes serve as two ''mirrors'', reflecting electron
into hole and hole into electron. For ballistic SNS junctions, discrete
Andreev bound states are formed in the N--region, each state carries
positive or negative supercurrent. For diffusive SNS junctions, the
so-called current carrying density of states (CCDOS) plays the similar role,
which also has positive and negative contributions to the supercurrent.
Recent experiment \cite{rev} demonstrated the $\pi $-junction transition in
the diffusive SNS junction by applying control voltage on the N-region. In
fact, the biased normal reservoirs across mesoscopic N--region induce a
non-equilibrium distribution in the N-region, and make the occupied fraction
of CCDOS deviating from the equilibrium one. When the control voltage is
excess a certain value, the non-equilibrium distribution has so much weight
on the negative part of CCDOS that the total current reverses its sign. Many
theoretical works have been addressed on this issue, either for ballistic
SNS junctions \cite{bll1,bll2,bll3}, or for diffusive SNS junctions \cite
{dff1,dff2,dff3,dff4}.

In addition to non-equilibrium distribution, there is a completely different
mechanism to realize the $\pi $-junction transition, i.e., coupling
superconductors by an Anderson impurity or an interacting quantum dot (QD).
The works of Glazman ${\sl et}$ ${\sl al}$. \cite{sds1} and Spivak ${\sl et}$
${\sl al}$. \cite{sds2} revealed that when the impurity is single occupied,
the sign of Josephson current for infinite Coulomb repulsion is opposite to
that without the repulsion. Ishizaka ${\sl et}$ ${\sl al}$. \cite{sds3}
obtained the condition for $\pi $-junction transition, by using non-crossing
approximation and varying the strength of the Coulomb repulsion, the bare
level position, the tunneling strength, and the temperature. Rozhkov ${\sl et%
}$ ${\sl al}$. \cite{sds4}\ analyzed the system in a non-perturbative way,
and found a novel intermediate phase in which one of $\phi =0$ and $\phi
=\pi $ is stable while the other is metastable, with the energy $E(\phi )$
having a kink somewhere in between. Clerk ${\sl et}$ ${\sl al}$. \cite{sds5}
studied the case of infinite-$U$ and the regime where the superconducting
gap $\Delta $ and the Kondo temperature $T_K$ are comparable, showing that
the position of the sub-gap resonance in the impurity spectral function
develops a strong anomalous phase dependence, and the $\pi $-junction
behavior is lost as the position of the bound state moves above the Fermi
energy.

Recently, there are growing interests in the physics of superconductor in
contact with ferromagnetic material, and the following works revealed
another approach to achieve $\pi $-junction. Proki\'{c} ${\sl et}$ ${\sl al}$%
. \cite{sfs1} presented a theory of the $\pi $-junction transition in
atomic-scale superconductor / ferromagnet (S/F) superlattices. They found
that the critical Josephson current has a non-monotonic dependence on the
exchange field $h$ in the ferromagnetic layer, becoming zero at the critical
value, corresponding to the transition between $\phi =0$ and $\phi =\pi $ in
the ground state. Yip ${\sl et}$ ${\sl al}$. \cite{sfs2} and Heikkil\"{a} $%
{\sl et}$ ${\sl al}$. \cite{sfs3} demonstrated that the supercurrent through
a mesoscopic SFS junction oscillates with an exponential decreasing envelope
as a function of the exchange field or the distance between the electrodes.
They also proposed that the suppressed supercurrent by the exchange field
can be recovered by a proper non-equilibrium distribution.

With these in mind, we are curious whether there is any relationship among
the above three mechanisms for $\pi $-junction transition. Motivated by this
question, we investigate the following cases of $\pi $-junction transition
in a superconductor / quantum dot / superconductor (S-QD-S) system and
provide a unified picture based on Andreev bound states. In section II, we
study the $\pi $-junction transition caused by Zeeman splitting. Assuming
the QD with two spin levels $E_{\uparrow }$ and $E_{\downarrow }$, we find
that the surface of the Josephson current $I(\frac \pi 2)$ vs the
configuration of $(E_{\uparrow },E_{\downarrow })$ exhibits interesting
profile: a sharp peak around $E_{\uparrow }=E_{\downarrow }=0$; a positive
ridge in the region of $E_{\uparrow }\cdot E_{\downarrow }>0$; and a ${\em %
negative}$, flat, shallow plain in the region of $E_{\uparrow }\cdot
E_{\downarrow }<0$. In section III, we study the $\pi $-junction transition
caused by intra-dot interaction. We model QD by $H_{dot}=E_0\sum_\sigma
c_\sigma ^{\dagger }c_\sigma +Un_{\uparrow }n_{\downarrow }$, and handle the
interaction term by Hartree-Fock approximation. Thus this case is reduced to
the first one except a self-consistent calculation for $\left\langle
n_\sigma \right\rangle $. We show that the $\pi $-junction transition occurs
when QD becomes a magnetic dot due to the interaction. The conditions for $%
\pi $-junction transition are also discussed. In section IV, we study the $%
\pi $-junction transition caused by non-equilibrium distribution in QD. By
replacing the Fermi distribution $f(\omega )$ with a non-equilibrium one $%
\frac 12\left[ f(\omega -V_c)+f(\omega +V_c)\right] $, and allowing Zeeman
splitting in QD as $E_{\uparrow }=-E_{\downarrow }=h,$ we find that for $h=0$%
, the supercurrent reverses its sign when control voltage $V_c$ is excess a
certain value, which is agree with previous work \cite{ar4} and the
experiment \cite{rev}. For $h\neq 0$, the curves of Josephson current vs
control voltage show a novel effect of interplay of non-equilibrium
distribution with the magnetization in QD. Finally, we summarize our
understanding of $\pi $-junction transition in S-QD-S system in section V.

\section{ QD WITH TWO SPIN LEVELS}

\subsection{model Hamiltonian and formulation}

In this section, we study the S-QD-S system modeled by the following
Hamiltonian:

\begin{eqnarray}
H &=&H_L+H_R+H_{dot}+H_T\;\;, \\
H_L &=&\sum_{k\sigma }\epsilon _ka_{k\sigma }^{\dagger }a_{k\sigma
}+\sum_k\left[ \Delta e^{-i\phi _L}a_{k\uparrow }^{\dagger }a_{-k\downarrow
}^{\dagger }+h.c\right] \;\;,  \nonumber \\
H_R &=&\sum_{p\sigma }\epsilon _pb_{p\sigma }^{\dagger }b_{p\sigma
}+\sum_p\left[ \Delta e^{-i\phi _R}b_{p\uparrow }^{\dagger }b_{-p\downarrow
}^{\dagger }+h.c\right] \;\;,  \nonumber \\
H_{dot} &=&\sum_\sigma E_\sigma c_\sigma ^{\dagger }c_\sigma \;\;,  \nonumber
\\
H_T &=&\sum_{k\sigma }\left[ t_La_{k\sigma }^{\dagger }c_\sigma +h.c\right]
+\sum_{p\sigma }\left[ t_Rb_{p\sigma }^{\dagger }c_\sigma +h.c\right] \;\;, 
\nonumber
\end{eqnarray}
where $H_L$, $H_R$ describe the left and right superconducting leads with
phase difference $\phi _L-\phi _R$ \cite{remark1}, $H_{dot}$ describes the
quantum dot with two spin levels, and $H_T$ is the coupling between the
quantum dot and the superconducting leads.

Since Josephson current can be expressed in terms of the Green functions of
the QD, we first derive the Green function by solving Dyson equation.
Following the formulation in \cite{ar3}, we denote ${\bf G}$ and ${\bf g}$
as the Green functions of QD in Nambu representation, with and without the
coupling to leads, respectively; and denote ${\bf \Sigma }$ as the
self-energy due to the coupling between QD\ and the leads. The retarded
Green function (Fourier transformed) of isolated QD is 
\begin{equation}
{\bf g}^r=\left( 
\begin{array}{cc}
\frac 1{\omega -E_{\uparrow }+\text{i}0^{+}} & 0 \\ 
0 & \frac 1{\omega +E_{\downarrow }+\text{i}0^{+}}
\end{array}
\right) \;\;.
\end{equation}
The retarded self-energy (Fourier transformed) under the wide bandwidth
approximation can be derived as \cite{ar3}, 
\begin{equation}
{\bf \Sigma }_{L/R}^r(\omega )=-\frac{\text{i}}2\Gamma _{L/R}\rho (\omega
)\left( 
\begin{array}{cc}
1 & -\frac \Delta \omega e^{-\text{i}\phi _{L/R}} \\ 
-\frac \Delta \omega e^{\text{i}\phi _{L/R}} & 1
\end{array}
\right) \;\;.
\end{equation}
where $\Gamma _{L/R}$ is the coupling strength between the superconducting
leads and QD, defined by $\Gamma _{L/R}\equiv 2\pi N_{L/R}t_{L/R}^2$, in
which $N_L$ and $N_R$ are the density of states in the left and right leads
in normal state. The factor $\rho (\omega )$ is defined as,

\begin{equation}
\rho (\omega )\equiv \left\{ 
\begin{array}{cc}
\frac{|\omega |}{\sqrt{\omega ^2-\Delta ^2}} & \;\;\;\;|\omega |>\Delta \\ 
\frac \omega {\text{i}\sqrt{\Delta ^2-\omega ^2}} & \;\;\;\;|\omega |<\Delta
\end{array}
\right. \;\;.
\end{equation}
Notice that $\rho (\omega )$ is the ordinary dimensionless BCS\ density of
states when $|\omega |>\Delta $, but has an imaginary part when $|\omega
|<\Delta $, corresponding to Andreev reflection process within the
superconducting gap. For simplicity, we assume that the two superconducting
leads are identical except a phase difference . Let $\phi _L=\frac \phi 2$, $%
\phi _R=-\frac \phi 2$, $\Gamma _L=\Gamma _R\equiv \Gamma $, then we obtain

\begin{eqnarray}
{\bf \Sigma }^r &\equiv &{\bf \Sigma }_L^r{\bf +\Sigma }_R^r=-\text{i}\Gamma
\rho (\omega )\left( 
\begin{array}{cc}
1 & -\frac \Delta \omega \cos \frac \phi 2 \\ 
-\frac \Delta \omega \cos \frac \phi 2 & 1
\end{array}
\right) \;\;, \\
{\bf \tilde{\Sigma}}^r &\equiv &{\bf \Sigma }_L^r{\bf -\Sigma }_R^r=-\text{i}%
\Gamma \rho (\omega )\left( 
\begin{array}{cc}
0 & -\frac \Delta \omega (-\text{i})\sin \frac \phi 2 \\ 
-\frac \Delta \omega \text{i\ }\sin \frac \phi 2 & 0
\end{array}
\right) \;\;.
\end{eqnarray}
By using Dyson equation, the retarded Green function of QD can be obtained
as,

\begin{equation}
{\bf G}^r{\bf =}\left[ {\bf g}^{r^{-1}}{\bf -\Sigma }^r\right] ^{-1}=\frac 1A%
\left( 
\begin{array}{cc}
g_{22}^{r^{-1}}-\Sigma _{22}^r & \Sigma _{12}^r \\ 
\Sigma _{21}^r & g_{11}^{r^{-1}}-\Sigma _{11}^r
\end{array}
\right) \;\;,
\end{equation}
where $A=A(\omega )$ defined as 
\begin{equation}
A(\omega )\equiv \det \left[ {\bf g}^{r^{-1}}{\bf -\Sigma }^r\right]
=(g_{22}^{r^{-1}}-\Sigma _{22}^r)(g_{11}^{r^{-1}}-\Sigma _{11}^r)-\Sigma
_{12}^r\;\Sigma _{21}^r\;\;.
\end{equation}

The general current formula for a mesoscopic hybrid multi-terminal system
has been derived in \cite{ar3}. For the time-independent case and QD with
two spin levels under consideration, the current formula can be rewritten in
a compact form (in units of $e=\hbar =1$): 
\begin{equation}
I_{L/R}=I_{L/R,\uparrow }+I_{L/R,\downarrow }=\int \frac{d\omega }{2\pi }2%
\mathop{\rm Re}%
\left[ {\bf G\Sigma }_{L/R}\right] _{11-22}^{<}\;\;,
\end{equation}
where $\left[ CD\right] ^{<}\equiv C^{<}D^a+C^rD^{<}$, $\left[ \;\;\right]
_{11-22}\equiv \left[ \;\;\right] _{11}-\left[ \;\;\right] _{22}$, and ${\bf %
G}$, ${\bf \Sigma }_{L/R}$ are the Fourier transformed $2\times 2$ Nambu
matrices. Since $I=I_L=-I_R$ in the stationary transport, the current
formula can be further reduced to

\begin{equation}
I=\frac 12(I_L-I_R)=\int \frac{d\omega }{2\pi }%
\mathop{\rm Re}%
\left[ {\bf G\tilde{\Sigma}}\right] _{11-22}^{<}\;\;,
\end{equation}
with ${\bf \tilde{\Sigma}\equiv \Sigma }_L{\bf -\Sigma }_R$ . Applying the
fluctuation-dissipation theorem, one has ${\bf G}^{<}{\bf =}f(\omega )\left[ 
{\bf G}^a{\bf -G}^r\right] $ and ${\bf \tilde{\Sigma}}^{<}=$ $f(\omega
)\left[ {\bf \tilde{\Sigma}}^a{\bf -\tilde{\Sigma}}^r\right] $, where $%
f(\omega )=1/(e^{\beta \omega }+1)$ is the Fermi distribution function.
Notice that $({\bf G}^r{\bf )}^{\dagger }{\bf =G}^a$, $({\bf \tilde{\Sigma}}%
^r{\bf )}^{\dagger }{\bf =\tilde{\Sigma}}^a$, the expression in the
integrand can be simplified to

\begin{equation}
\mathop{\rm Re}%
\left[ {\bf G\tilde{\Sigma}}\right] _{11-22}^{<}=f(\omega )2\sin \phi \frac{%
\Gamma ^2\Delta ^2}{\omega ^2-\Delta ^2}\left[ -%
\mathop{\rm Im}%
\frac 1{A(\omega )}\right] \;\;.
\end{equation}
Consequently, the Josephson current is expressed as 
\begin{equation}
I=2\sin \phi \int \frac{d\omega }{2\pi }f(\omega )j(\omega )\;\;,
\end{equation}
in which the current carrying density of states (CCDOS) $j(\omega )$ is
defined by 
\begin{equation}
j(\omega )\equiv \frac{\Gamma ^2\Delta ^2}{\omega ^2-\Delta ^2}\left[ -%
\mathop{\rm Im}%
\frac 1{A(\omega )}\right] \;\;.
\end{equation}
Because the singularities of $j(\omega )$ lie in the same half-plain, CCDOS $%
j(\omega )$ satisfies the condition $\int j(\omega )d\omega =0$.

Since $\Sigma ^r(\omega )$ is purely imaginary when $\left| \omega \right|
>\Delta $ while purely real when $\left| \omega \right| <\Delta $, so $%
A(\omega )$ has finite imaginary part when $\left| \omega \right| >\Delta $,
while infinitesimal imaginary part when $\left| \omega \right| <\Delta $.
Correspondingly, the Josephson current can be divided into two parts,
contributed from the continuous spectrum and from the discrete spectrum,
respectively:

\begin{eqnarray}
I &=&I_c+I_d\;\;, \\
I_c &\equiv &2\sin \phi \left( \int\nolimits_{-\infty }^{-\Delta
}+\int\nolimits_\Delta ^\infty \right) \frac{d\omega }{2\pi }f(\omega
)j(\omega )\;\;,  \nonumber \\
I_d &\equiv &2\sin \phi \int\nolimits_{-\Delta }^\Delta \frac{d\omega }{2\pi 
}f(\omega )j(\omega )\;\;.  \nonumber
\end{eqnarray}
We shall show in the appendix that when $|\omega |<\Delta $ and $\phi \neq 0$%
, the equation $A(\omega )=0$ has two real roots denoted by $\tilde{E}%
_1=\Delta \sin \theta _1$and $\tilde{E}_2=\Delta \sin \theta _2$, where $%
\theta _1$ and $\theta _2$ are the two roots of the following equation, 
\begin{equation}
\left( \sin \theta +\frac \Gamma \Delta \tan \theta -\frac{E_{\uparrow }}%
\Delta \right) \left( \sin \theta +\frac \Gamma \Delta \tan \theta +\frac{%
E_{\downarrow }}\Delta \right) \cos ^2\theta -\frac{\Gamma ^2}{\Delta ^2}%
\cos ^2\frac \phi 2=0
\end{equation}
with $\theta \in (-\frac \pi 2,\frac \pi 2)$. Equation (15) completely
determines the properties of Andreev bound states. The roots $\tilde{E}_1$%
and $\tilde{E}_2$ are just the Andreev bound states. In the range of $\left|
\omega \right| <\Delta $, $A(\omega )$ can be written as $a(\omega )(\omega -%
\tilde{E}_1+$i$0^{+})(\omega -\tilde{E}_2+$i$0^{+})$, and $\left[ -%
\mathop{\rm Im}%
\frac 1{A(\omega )}\right] $ is reduced to 
\begin{equation}
-%
\mathop{\rm Im}%
\frac 1{A(\omega )}=\pi \left[ \frac 1{A^{\prime }(\tilde{E}_1)}\delta
(\omega -\tilde{E}_1)+\frac 1{A^{\prime }(\tilde{E}_2)}\delta (\omega -%
\tilde{E}_2)\right] \;\;.  \nonumber
\end{equation}
Finally, the Josephson current through S-QD-S system can be expressed as, 
\begin{eqnarray}
I &=&I_c+I_d\;\;, \\
I_c &\equiv &2\sin \phi \left( \int\nolimits_{-\infty }^{-\Delta
}+\int\nolimits_\Delta ^\infty \right) \frac{d\omega }{2\pi }f(\omega )\frac{%
\Gamma ^2\Delta ^2}{\omega ^2-\Delta ^2}\left[ -%
\mathop{\rm Im}%
\frac 1{A(\omega )}\right] \;\;,  \nonumber \\
I_d &\equiv &\sin \phi \left[ f(\tilde{E}_1)\frac{\Gamma ^2\Delta ^2}{\tilde{%
E}_1^2-\Delta ^2}\frac 1{A^{^{\prime }}(\tilde{E}_1)}+f(\tilde{E}_2)\frac{%
\Gamma ^2\Delta ^2}{\tilde{E}_2^2-\Delta ^2}\frac 1{A^{^{\prime }}(\tilde{E}%
_2)}\right] \;\;.  \nonumber
\end{eqnarray}
This current formula will be used in the following numerical study \cite
{remark2}.

\subsection{numerical results and discussions}

Now we discuss the numerical results for Andreev bound states and the
Josephson current. In all numerical studies of this paper, we take $e=\hbar
=k_B=1$, set $\Delta =1$, i.e., measure all energies in units of $\Delta $,
let $\Gamma =0.1$ for symmetric and weak coupling case, and fix the phase
difference $\phi =\frac \pi 2$.

Fig.1 presents the solution of Eq.(15), i.e., Andreev bound states of S-QD-S
system. In the limit of $\Gamma \rightarrow \infty $, Eq.(15) gives $\tilde{E%
}_1=\left| \cos \frac \phi 2\right| $ and $\tilde{E}_2=-\left| \cos \frac %
\phi 2\right| $, the well-known Andreev bound states for a clean
superconducting point contact \cite{abs}. Conversely, in the limit of $%
\Gamma \rightarrow 0$, the two roots of Eq.(15) are $\tilde{E}_1=E_{\uparrow
}$ and $\tilde{E}_1=-E_{\downarrow }$, i.e., the bare levels of electron
with spin$\uparrow $ and hole with spin$\downarrow $ of the QD. (We describe
spin$\uparrow $ quasiparticle in electron language and spin$\downarrow $
quasiparticle in hole language due to the choice of Nambu representation.)
For the case of $\Gamma \ll \Delta $ under consideration, where QD is weakly
coupled with the superconducting leads, the solutions of Eq.(15) depend
strongly on the configuration of QD levels $(E_{\uparrow },E_{\downarrow })$%
, but weakly on the phase difference $\phi $. The surfaces of $\tilde{E}_1$
vs $(E_{\uparrow },E_{\downarrow })$ and $\tilde{E}_2$ vs $(E_{\uparrow
},E_{\downarrow })$ for $\phi =\frac \pi 2$ are shown in Fig.1a and Fig.1b,
respectively. In this case, the electron level $E_{\uparrow }$ and the hole
level $-E_{\downarrow }$ are coupled by AR\ tunneling. Therefore, Andreev
bound states can be viewed as hybrid states of $E_{\uparrow }$ and $%
-E_{\downarrow }$, and an energy gap of the order $\Gamma $ is opened where $%
E_{\uparrow }$ and $-E_{\downarrow }$ are equal (see Fig.1c and Fig.1d).
Further study on the relations of $\tilde{E}_1$ vs $\phi $ and $\tilde{E}_2$
vs $\phi $ can provide the information of supercurrent carried by each
Andreev bound state (not shown).

Fig.2 presents the surface of Josephson current $I$ vs the configuration of $%
(E_{\uparrow },E_{\downarrow })$. First, the surface is symmetric to the
diagonal lines of $E_{\uparrow }=E_{\downarrow }$ and $E_{\uparrow
}=-E_{\downarrow }$, which reflects the symmetry between electron and hole
and the fact that supercurrent are non-spin-polarized. Second, the surface
exhibits interesting profile: a sharp peak around $E_{\uparrow
}=E_{\downarrow }=0$; a positive ridge in the region of $E_{\uparrow }\cdot
E_{\downarrow }>0$; and a negative, flat, shallow plain in the region of $%
E_{\uparrow }\cdot E_{\downarrow }<0$. Note that the positive ridge and
negative plain share a sharp edge. Third, consider two special but typical
cases, $E_{\uparrow }=E_{\downarrow }\equiv E_0$ (Fig.2b), and $E_{\uparrow
}=-E_{\downarrow }\equiv E_0$ (Fig.2c), which are actually the diagonal cuts
of Fig.2a. For the case of $E_{\uparrow }=E_{\downarrow }$, the $I$ vs $E_0$
curve has a peak at $E_0=0$ with a width of $\Gamma $, which reproduces the
result for QD with one spin-degenerate level \cite{ar4}. For the case of $%
E_{\uparrow }=-E_{\downarrow }$, the $I$ vs $E_0$ curve has the same maximum
at $E_0=0$, but jumps suddenly from the positive maximum to a small negative
value around $E_0=\pm \Gamma $, which is quite similar to the curve of the
critical current vs the exchange field in an atomic scale S/F superlattice
(see Fig.2 of \cite{sfs1}).

To understand these features, we plot CCDOS $j(\omega )$ for different
configurations of $(E_{\uparrow },E_{\downarrow })$ in Fig.3. The
supercurrent can be expressed as $I=2\sin \phi \int \frac{d\omega }{2\pi }%
f(\omega )j(\omega )$, and only the spectrum of $\omega <0$ devotes to the
current at zero temperature. For the spectrum of $E_{\uparrow
}=E_{\downarrow }=0$ , $j(\omega )$ has two $\delta $-function type discrete
spectrum within the superconducting gap, corresponding to two Andreev bound
states. They carry supercurrent with opposite signs, positive for $\tilde{E}%
_1<0$, and negative for $\tilde{E}_2>0$. $j(\omega )$ also has continuous
spectrum outside the superconducting gap, negative and positive for $\omega
<-\Delta $ and $\omega >\Delta $, respectively. Since the contribution from
the discrete spectrum is much larger than that from the continuous one, the
current peak at $E_{\uparrow }=E_{\downarrow }=0$ is mostly contributed from 
$\tilde{E}_1$. For the spectrum of $E_{\uparrow }=E_{\downarrow }=E_0\neq 0$%
, $\tilde{E}_1$ and $\tilde{E}_2$ move toward $\pm \Delta $, symmetric to
the Fermi surface (see also Fig.1c). The contribution from the discrete
spectrum $\tilde{E}_1$ decreases continuously with $E_0$, corresponding to
the $\Gamma $-width broadening of the $E_0=0$ peak in Fig.2c. For the
spectrum of $E_{\uparrow }=-E_{\downarrow }=E_0\neq 0$, however, $\tilde{E}%
_1 $ and $\tilde{E}_2$ move in the same direction. When $E_0<-\Gamma $ or $%
E_0>\Gamma $, both $\tilde{E}_1$ and $\tilde{E}_2$ are below or above the
Fermi surface (see also Fig.1d). As a consequence, they have little net
contribution to the supercurrent, and the relatively small negative
continuous spectrum of $\omega <-\Delta $ dominates. Because the crossover
of $\tilde{E}_1$ and $\tilde{E}_2$ from different sides of the Fermi surface
to one side occurs abruptly, sudden jumps between the positive maximum and
the negative valleys appear in Fig.2d. Similarly, one can understand the
whole surface of Fig.2 with the help of the Andreev bound states in Fig.1
and the properties of CCDOS in Fig.3.

The above results are for zero temperature, the temperature dependence of
the Josephson current is shown in Fig.4. Notice that the supercurrent is
sensitive to the temperature. The height of $E_{\uparrow }=E_{\downarrow }=0$
peak (referred as $I_0$) decreases rapidly with the increase of the
temperature, $I_0=0.06,\;0.004,\;0.001$ for $T=0,\;0.25,\;0.50$,
respectively. The sharp edge between $I>0$ and $I<0$ is also smeared out at
finite temperatures. This reason is, at finite temperature not only the
Andreev bound state below the Fermi surface but also the one above the Fermi
surface has contribution to the supercurrent.

\section{QD WITH INTRA-DOT INTERACTION}

\subsection{Hartree-Fock approximation}

In this section, we investigate the $\pi $-junction transition caused by the
intra-dot interaction. The quantum dot with Coulomb interaction can be
described by $H_{dot}=E_0\sum_\sigma c_\sigma ^{\dagger }c_\sigma
+Un_{\uparrow }n_{\downarrow }$. As in the problem of local moment in
nonmagnetic metals \cite{mag}, we deal with the interaction term by
Hartree-Fock approximation (HFA), in which $Un_{\uparrow }n_{\downarrow }$
is replaced by $U\left\langle n_{\uparrow }\right\rangle n_{\downarrow
}+Un_{\uparrow }\left\langle n_{\downarrow }\right\rangle $. Thus, $H_{dot}$
becomes $\sum_\sigma E_\sigma ^{\prime }c_\sigma ^{\dagger }c_\sigma ,$ with
the effective levels $E_\sigma ^{\prime }$ $\equiv E_0+U\left\langle n_{\bar{%
\sigma}}\right\rangle $. Despite of the roughness of HFA, it contains the
physics of magnetization due to Coulomb interaction. More important, the
approximation allows us to obtain a solution including infinite order of
tunneling processes, which is crucial for describing the Andreev bound
states and the supercurrent they carried. However, HFA fails in the Kondo
regime. As discussed in \cite{sds5}, if $\Delta \ll T_K\equiv \sqrt{U\Gamma }%
e^{-\pi |E_0-\mu |/\Gamma }$, the spin of the impurity is completely
screened by Kondo effect, and there will be no $\pi $-junction behavior. In
this work, we will constrain ourselves to the weak coupling case where
magnetization effect dominates, corresponding to $\Delta \gg T_K$ regime.

Then most of the formula in the previous section can be transplanted, except
the averaged occupation number $\left\langle n_\sigma \right\rangle $ needs
a self-consistent calculation. Notice that $\left\langle n_\sigma
\right\rangle $ can be derived from the retarded Green function of QD as 
\begin{eqnarray}
\left\langle n_{\uparrow }\right\rangle &=&\int \frac{d\omega }{2\pi }%
f(\omega )\left[ -2%
\mathop{\rm Im}%
G_{11}^r(\omega )\right] \;\;, \\
1-\left\langle n_{\downarrow }\right\rangle &=&\int \frac{d\omega }{2\pi }%
f(\omega )\left[ -2%
\mathop{\rm Im}%
G_{22}^r(\omega )\right] \;\;.
\end{eqnarray}
Similar to the current formula, $\left\langle n_\sigma \right\rangle $ can
be divided into two parts, the contribution from discrete spectrum and the
contribution from continuous spectrum, 
\begin{eqnarray}
\left\langle n_{\uparrow }\right\rangle &=&\left\langle n_{\uparrow
}\right\rangle _c+\left\langle n_{\uparrow }\right\rangle _d\;\;, \\
\left\langle n_{\uparrow }\right\rangle _c &=&\left( \int\nolimits_{-\infty
}^{-\Delta }+\int\nolimits_\Delta ^\infty \right) \frac{d\omega }{2\pi }%
f(\omega )\left[ -2%
\mathop{\rm Im}%
\frac{g_{22}^{r^{-1}}(\omega )-\Sigma _{22}^r(\omega )}{A(\omega )}\right]
\;\;,  \nonumber \\
\left\langle n_{\uparrow }\right\rangle _d &=&\sum_{i=1}^2\left[ f(\omega )%
\frac 1{A^{^{\prime }}(\omega )}\left( g_{22}^{r^{-1}}(\omega )-\Sigma
_{22}^r(\omega )\right) \right] _{\omega =\tilde{E}_i}\;\;,  \nonumber \\
\left\langle n_{\downarrow }\right\rangle &=&\left\langle n_{\downarrow
}\right\rangle _c+\left\langle n_{\downarrow }\right\rangle _d\;\;, \\
1-\left\langle n_{\downarrow }\right\rangle _c &=&\left(
\int\nolimits_{-\infty }^{-\Delta }+\int\nolimits_\Delta ^\infty \right) 
\frac{d\omega }{2\pi }f(\omega )\left[ -2%
\mathop{\rm Im}%
\frac{g_{11}^{r^{-1}}(\omega )-\Sigma _{11}^r(\omega )}{A(\omega )}\right]
\;\;,  \nonumber \\
1-\left\langle n_{\downarrow }\right\rangle _d &=&\sum_{i=1}^2\left[
f(\omega )\frac 1{A^{^{\prime }}(\omega )}\left( g_{11}^{r^{-1}}(\omega
)-\Sigma _{11}^r(\omega )\right) \right] _{\omega =\tilde{E}_i}\;\;. 
\nonumber
\end{eqnarray}
Since ${\bf g}^r$ contains unknown quantity $\left\langle n_\sigma
\right\rangle $ through $E_\sigma ^{\prime }$, the above equations for $%
\left\langle n_\sigma \right\rangle $ should be solved self-consistently.

\subsection{numerical results and discussions}

Next, we present the numerical results for the interacting QD system at zero
temperature. Fig.5 shows $I$ vs $E_0$ and corresponding $\left\langle
n_\sigma \right\rangle $ vs $E_0$ curves for two typical cases, the QD
without interaction ($U=0$ in Fig.5a) and with strong interaction ($U\gg
\Gamma $ in Fig.5b). The S-QD-S system behaves like one of the above two
cases, Depending on the magnitudes of the interacting constant $U$ and the
coupling strength $\Gamma $, the S-QD-S system may behave differently. If $%
U<\Gamma $, the occupation numbers $\left\langle n_{\uparrow }\right\rangle $
and $\left\langle n_{\downarrow }\right\rangle $ are almost equal, leading
to the supercurrent always positive with a maximum at $E_0=-U/2$, and the
system behaves as same as the non-interacting one. On the contrast, if $%
U>\Gamma $, a symmetry breaking solution of Eq.(19) and Eq.(20) is energy
preferred, in which $\left\langle n_{\uparrow }\right\rangle $ and $%
\left\langle n_{\downarrow }\right\rangle $ are unequal, and QD\ becomes a
magnetic dot. Consequently, the effective level $(E_{\uparrow }^{\prime
},E_{\downarrow }^{\prime })$ occupy a series of configurations in the
negative plain of Fig.1, so the Josephson current has a small but negative
valley in $I$ vs $E_0$ curve. We determine the transition border of the
above two cases by the parametric diagram of $I$ vs $(-E_0/\Gamma ,\Gamma
/U) $ in Fig.6. By virtue of electron-hole symmetry, the diagram is
symmetric to $-E_0/U=0.5$. The black area in Fig.6b indicates the range of
parameters where the S-QD-S system behaves as a $\pi $-junction. One can see
from the diagram that the system is likely to transfer to $\pi $-junction
around $E_0=-U/2$.

The temperature dependence are studied in Fig.7, which shows $I$ vs $E_0$
curves with $U=1$ for different temperatures. Due to the electron-hole
symmetry, only half of the plot is shown. The sharp structure at zero
temperature is smoothed at finite temperatures, and the negative part of the
supercurrent vanishes above a critical temperature. These features can be
understood by taking account of the temperature effect on the supercurrent
through QD with two spin levels in Fig.4 and the temperature effect on the
averaged QD occupation numbers. We also study the diagram of $I$ vs $(U,T)$
where $E_0$ is set to $-U/2$ (not shown here). The transition line from $I<0$
to $I>0$ can be fitted as $U=0.17+7.5T$, which is consistent with the result
derived by non-crossing approximation in \cite{sds3}.

\section{QD IN NON-EQUILIBRIUM DISTRIBUTION}

Now we turn to discuss $\pi $-junction transition caused by non-equilibrium
distribution in QD. In a recent work by Sun ${\sl et}$ ${\sl al}$. \cite{ar4}%
, a mesoscopic four-terminal Josephson junction (S-QD-S with two normal
leads connected to QD) was studied. By using non-equilibrium Green function
method, they found that the supercurrent between the two superconducting
electrodes can be suppressed and even reversed by changing the dc voltage
applied across the two normal terminals. Here we only take the essential
point of that work but omit its tedious calculation, by simply assuming that
QD has a two-step distribution function as 
\begin{equation}
F(\omega )=\frac 12\left[ f(\omega -V_c)+f(\omega +V_c)\right] \stackrel{%
T\rightarrow 0}{\longrightarrow }\left\{ 
\begin{array}{ll}
1\;\;\; & \omega <-V_c \\ 
0.5 & -V_c<\omega <V_c \\ 
0 & \omega >V_c
\end{array}
\right. \;\;,
\end{equation}
corresponding to the limit of $\Gamma _2=\Gamma _4\rightarrow 0$ in \cite
{ar4}. Different from \cite{ar4}, here we allow the two spin levels of QD
have a Zeeman splitting, i.e., $E_{\uparrow }=-E_{\downarrow }=h$. The
curves of the Josephson current $I(\phi =\frac \pi 2)$ vs the control
voltage $V_c$ for different Zeeman splitting $h$ are shown in Fig.8. These
curves are step-like because $F(\omega )$ is step-like at $T=0$, and $%
j(\omega )$ has $\delta $-function type discrete spectrum in the range of $%
-\Delta <\omega <\Delta $. Either finite temperature or small broadening of
Andreev bound states will smooth the curves.

For $h=0$, $I$ reverses its sign around $V_c=\Gamma $, and the magnitude of
positive current is much larger than that of negative one, which is agree
qualitatively with the experiment \cite{rev} and previous work \cite{ar4}.
For $h\neq 0$, curves of $I$ vs $V_c$ have a peak around $V_c=h$, with the
width about $2\Gamma $, the height about half of that for $h=0$. On each
side of the peak, there is a negative current valley, where the system
behaves as $\pi $-junction. These results are also consistent with the
calculations for non-equilibrium SFS junction \cite{sfs2,sfs3}. One can
understand these curves by considering the two-step distribution $F(\omega )$
and CCDOS in Fig.4. For example, for the curve of $h=0.2$ (marked with ``1''
in Fig.8), both positive and negative Andreev\ bound states are above the
Fermi surface [see CCDOS of $(E_{\uparrow },E_{\downarrow })=(0.2,-0.2)$ ],
located near $h-\Gamma $ and $h+\Gamma ,$ respectively. When $V_c=h$, the
positive bound state has a weight of 0.5, while the negative one has 0,
reducing the peak height to one half of that for $h=0$. For both $%
V_c<h-\Gamma $ or $V_c>h+\Gamma $, the two Andreev bound states have the
same weights (either 0 or 0.5), and have little net contribution to the
supercurrent. Therefore, the negative continuous spectrum of $\omega
<-\Delta $ is dominant in the supercurrent, leading to the $\pi $-junction
transition on both sides of the positive peak.

\section{ CONCLUSIONS}

In this work, we have investigated different mechanisms for the $\pi $%
-junction transition in S-QD-S system. From the current formula $I=2\sin
\phi \int \frac{d\omega }{2\pi }F(\omega )j(\omega )$, one can see that the $%
\pi $-junction transition may originate from the change of CCDOS $j(\omega )$%
, or the change of the distribution function $F(\omega )$, or from the
changes of both. The two mechanisms discussed in sections II and III, the
Zeeman splitting and intra-dot interaction, are involved the change of CCDOS 
$j(\omega )$ only. These two mechanisms are closely connected since the
intra-dot interaction may induce magnetization in QD if the interaction is
strong enough. The third mechanism studied in section IV involves the change
of the distribution function in QD, and the interplay of the magnetization
with the non-equilibrium distribution in QD. It is interesting that the
change of CCDOS and the change of distribution have the similar effect on
the $\pi $-junction transition, where the positive and negative Andreev
bound states cancels each other, leaving the negative continuous spectrum
dominant in the supercurrent. The interplay of the two mechanisms lead to
the novel effect that the supercurrent suppressed by magnetization of the QD
can be partially recovered by a proper non-equilibrium distribution of
electrons in the QD.

\section*{ACKNOWLEDGMENTS}

This project was supported by NSFC under grant No.10074001. One of the
authors (T.-H. Lin) would also like to thank the support from the Visiting
Scholar Foundation of State Key Laboratory for Mesoscopic Physics in Peking
University.

\smallskip $^{*}$ To whom correspondence should be addressed.

\section{APPENDIX}

In this appendix, we discuss the equation of Andreev bound states, i.e. $%
A(\omega )=0$ with $|\omega |<\Delta $ : 
\begin{equation}
A(\omega )=\left( \omega -E_{\uparrow }+\frac{\Gamma \omega }{\sqrt{\Delta
^2-\omega ^2}}\right) \left( \omega +E_{\downarrow }+\frac{\Gamma \omega }{%
\sqrt{\Delta ^2-\omega ^2}}\right) -\frac{\Gamma ^2\Delta ^2}{\Delta
^2-\omega ^2}\cos ^2\frac \phi 2=0\;\;.  \eqnum{(A1)}
\end{equation}
Let $\Gamma =\gamma \Delta $ , $E_{\uparrow }=\varepsilon _1\Delta $ , $%
E_{\downarrow }=-\varepsilon _2\Delta $ , $\omega =\Delta \sin \theta $ , $%
\theta \in (-\frac \pi 2,\frac \pi 2)$, the equation becomes a dimensionless
form, 
\begin{equation}
b(\theta )\equiv (\sin \theta +\gamma \tan \theta -\varepsilon _1)(\sin
\theta +\gamma \tan \theta -\varepsilon _2)\cos ^2\theta =\gamma ^2\cos ^2%
\frac \phi 2\;\;.  \eqnum{(A2)}
\end{equation}

Note that the function of $y=\sin \theta +\gamma \tan \theta $ projects $%
\theta \in (-\frac \pi 2,\frac \pi 2)$ monotonously to $y\in (-\infty
,+\infty )$. One can find $\Theta _1$and $\Theta _2$ in $(-\frac \pi 2,\frac %
\pi 2)$, satisfying $\sin \Theta _1+\gamma \tan \Theta _1=\varepsilon _1$,
and $\sin \Theta _2+\gamma \tan \Theta _2=\varepsilon _2$. Suppose $\Theta
_1<\Theta _2$, we have $b(\theta )\geqslant 0$ for $\theta \in (-\frac \pi 2%
,\Theta _1]\cup [\Theta _2,\frac \pi 2)$, and $b(\theta )<0$ for $\theta \in
(\Theta _1,\Theta _2)$. Because $b\left( \pm \frac \pi 2\right) =\gamma ^2$,
Eq.(A2) has at least two roots $\theta _1\in $$(-\frac \pi 2,\Theta _1]$and $%
\theta _2\in [\Theta _2,\frac \pi 2)$ for $\phi \neq 0$. It is
straightforward to find the two roots by dichotomy method.

Next, we prove $b^{\prime }(\theta )\geqslant 0$ for $0<b(\theta )<\gamma ^2$
and $\theta \in [\Theta _2,\frac \pi 2)$, while $b^{\prime }(\theta
)\leqslant 0$ for $0<b(\theta )<\gamma ^2$ and $\theta \in (-\frac \pi 2%
,\Theta _1]$, so that $\theta _1$and $\theta _2$ are the only two roots in $%
(-\frac \pi 2,\Theta _1]$ and $[\Theta _2,\frac \pi 2)$. For $\theta \in
[\Theta _2,\frac \pi 2)$, $\sin \theta +\gamma \tan \theta -\varepsilon
_1\geqslant 0$, and $\sin \theta +\gamma \tan \theta -\varepsilon
_2\geqslant 0$. Define $x\equiv \left[ (\sin \theta +\gamma \tan \theta
-\varepsilon _1)(\sin \theta +\gamma \tan \theta -\varepsilon _2)\right]
^{1/2}$, because $0<b(\theta )<\gamma ^2,$so one obviously has $0<x<\gamma
\sec \theta $. 
\begin{eqnarray}
b^{\prime }(\theta ) &=&2(\sin \theta +\gamma \tan \theta -\frac{\varepsilon
_1+\varepsilon _2}2)(\gamma +\cos ^3\theta )  \nonumber \\
&&-2\sin \theta \cos \theta (\sin \theta +\gamma \tan \theta -\varepsilon
_1)(\sin \theta +\gamma \tan \theta -\varepsilon _2)  \eqnum{(A3)} \\
&\geqslant &2x\gamma -2\cos \theta \cdot x^2=-2\cos \theta \cdot x(x-\gamma
\sec \theta )\geqslant 0\;\;.  \nonumber
\end{eqnarray}
Therefore, Eq.(A2) has and only has two roots in the range of $\theta \in (-%
\frac \pi 2,\frac \pi 2)$ for $\phi \neq 0$.

For $\phi =0$, because $b^{\prime }\left( \frac \pi 2\right) =-2\gamma
[2-(\varepsilon _1+\varepsilon _2)]$, $b^{\prime }\left( -\frac \pi 2\right)
=2\gamma [2+(\varepsilon _1+\varepsilon _2)]$, and considering the
properties of $b^{\prime }(\theta )$, one can clearly see that Eq.(A2) has
two roots in $(-\frac \pi 2,\frac \pi 2)$if $|\varepsilon _1+\varepsilon
_2|<2$, but only has one root in $(-\frac \pi 2,\frac \pi 2)$if $%
|\varepsilon _1+\varepsilon _2|\geqslant 2$. However, this case is
irrelevant to the Josephson effect.


\newpage

\section*{Figure Captions}

\begin{itemize}
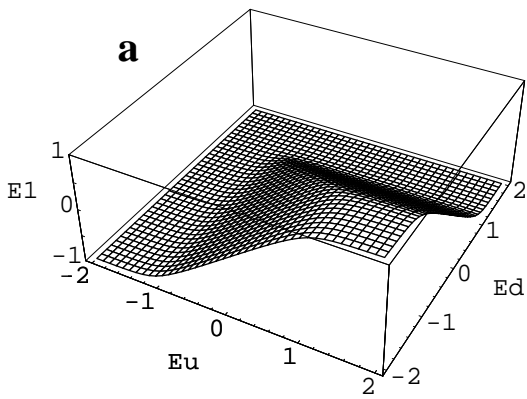
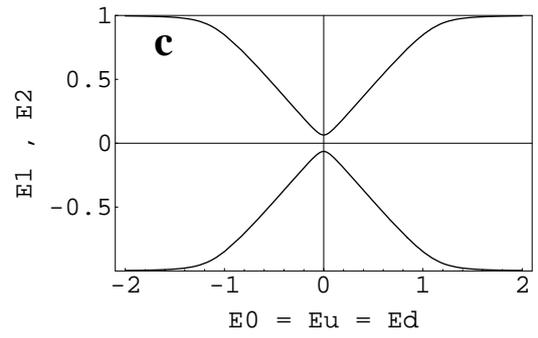
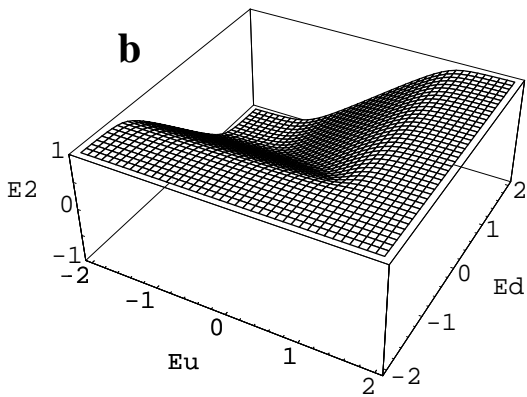
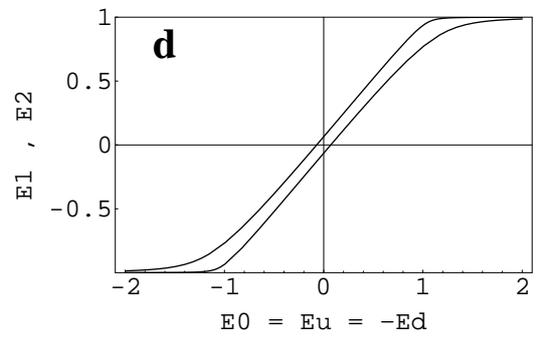
\item[{\bf Fig. 1}]  Solution of the equation for Andreev bound states.
Parameters are: $\Delta =1,\;\Gamma =0.1,\;\phi =\frac \pi 2$. (a) and (b)
show the two roots $\tilde{E}_1$ and $\tilde{E}_2$ ($\tilde{E}_1>\tilde{E}_2$%
) vs the configurations of QD levels $(E_{\uparrow },E_{\downarrow })$. (c)
and (d) are the diagonal cuts of (a) and (b), showing $\tilde{E}_1$ and $%
\tilde{E}_2$ vs $E_0$ with $E_{\uparrow }=E_{\downarrow }\equiv E_0$ and $%
E_{\uparrow }=-E_{\downarrow }\equiv E_0$, respectively. $\tilde{E}_1$ and $%
\tilde{E}_2$ can be viewed as two hybrid levels of the electron level of $%
E=E_{\uparrow }$ and the hole level of $E=-E_{\downarrow }$. ($\tilde{E}_1$, 
$\tilde{E}_2$, $E_{\uparrow }$ and $E_{\downarrow }$ are marked as E1, E2,
Eu and Ed in the plot, respectively.)

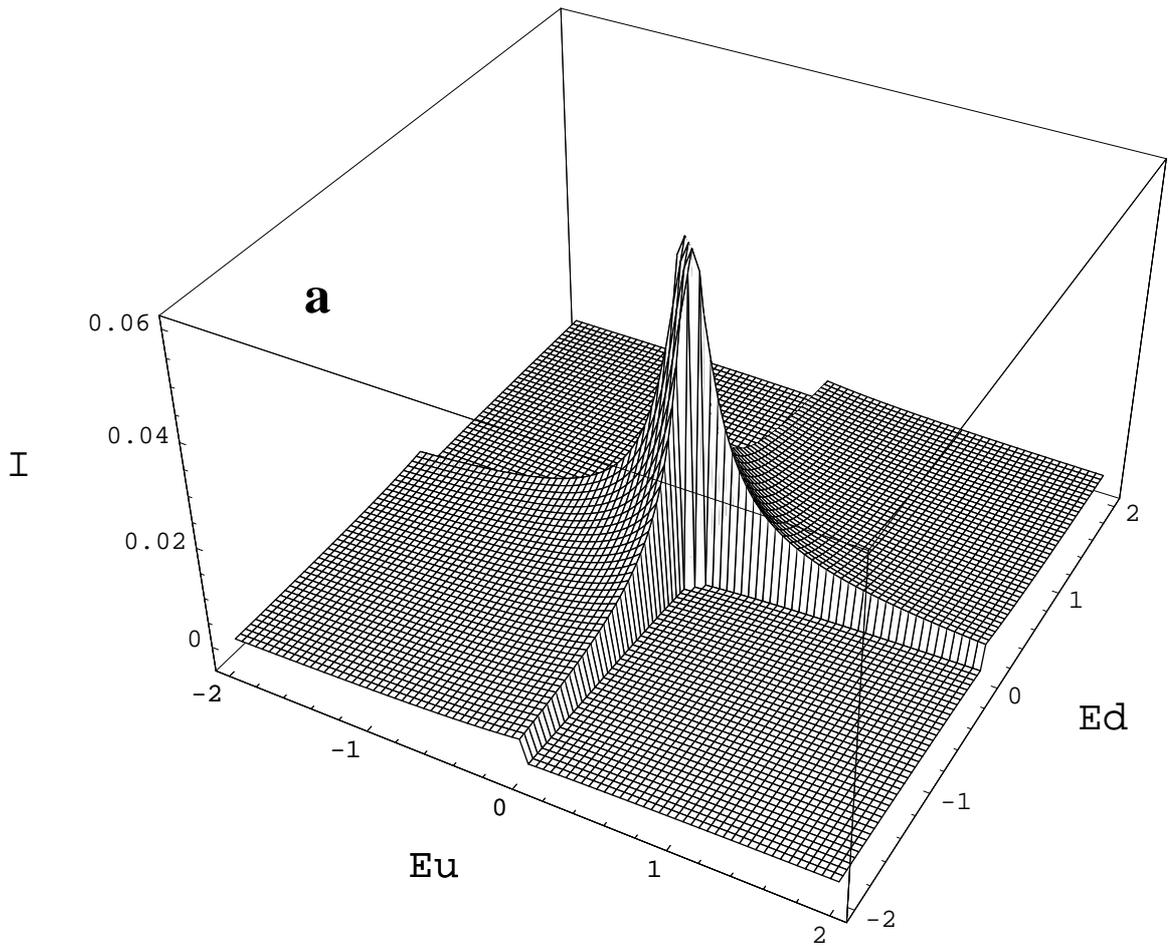
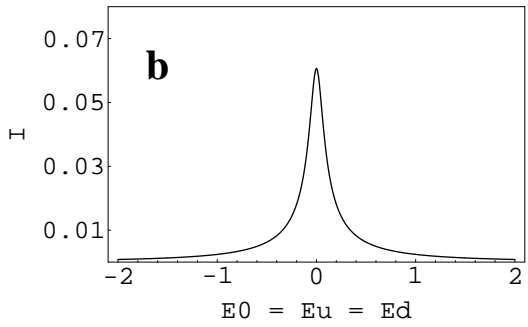
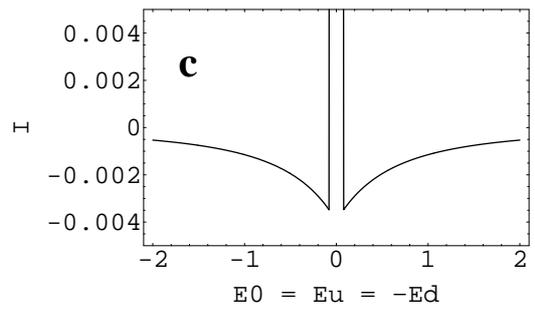
\item[{\bf Fig. 2}]  Surface of the Josephson current $I$ vs the
configuration of QD levels $(E_{\uparrow },E_{\downarrow })$. Parameters
are: $\Delta =1,\;\Gamma =0.1,\;\phi =\frac \pi 2,\;T=0$. (a) is the surface
graph, while (b), (c) are the diagonal cuts. (b) and (c) show $I$ vs $E_0$
with $E_{\uparrow }=E_{\downarrow }\equiv E_0$ and $E_{\uparrow
}=-E_{\downarrow }\equiv E_0$, respectively.

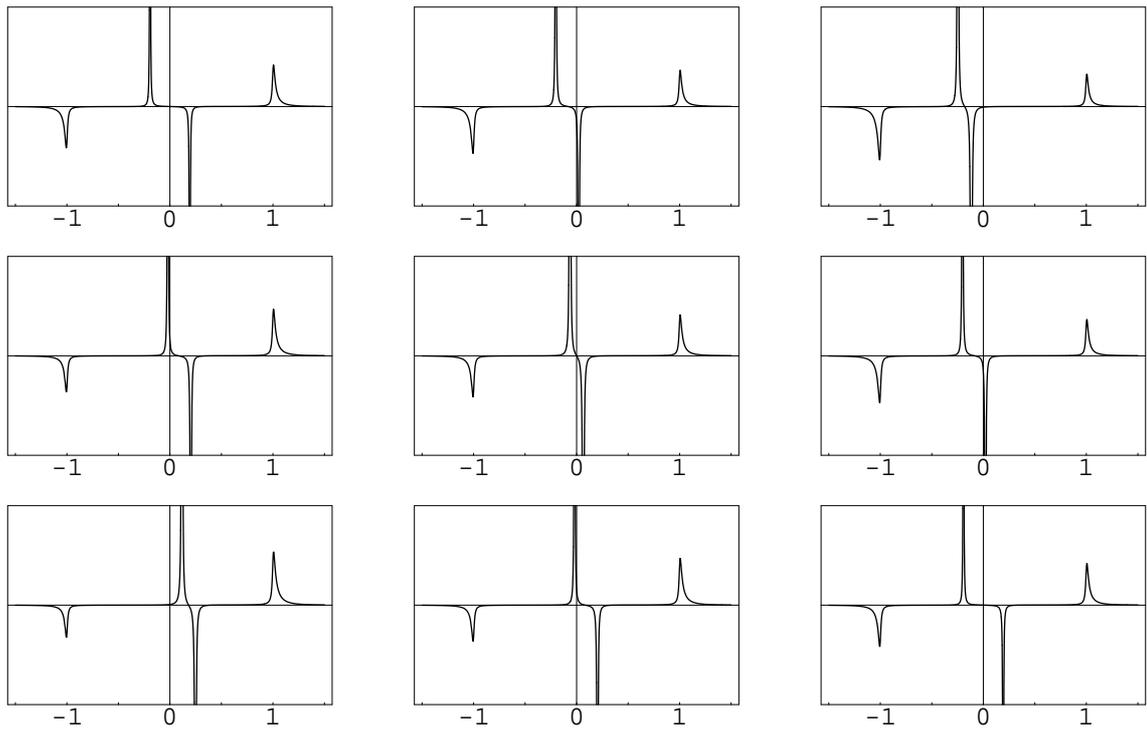
\item[{\bf Fig. 3}]  CCDOS $j(\omega )$ for different $(E_{\uparrow
},E_{\downarrow })$ configurations. The row from up to down corresponds to $%
E_{\uparrow }=-0.2,\;0,\;0.2$, and the column from left to right corresponds
to $E_{\downarrow }=-0.2,\;0,\;0.2$. Each of CCDOS contains two types of
spectrum: the discrete spectrum in the range of $|\omega |<1$ and the
continuous spectrum in the range of $|\omega |>1$. To illustrate discrete
spectrum, we broaden $\delta $ functions by 0.01 in the plots.

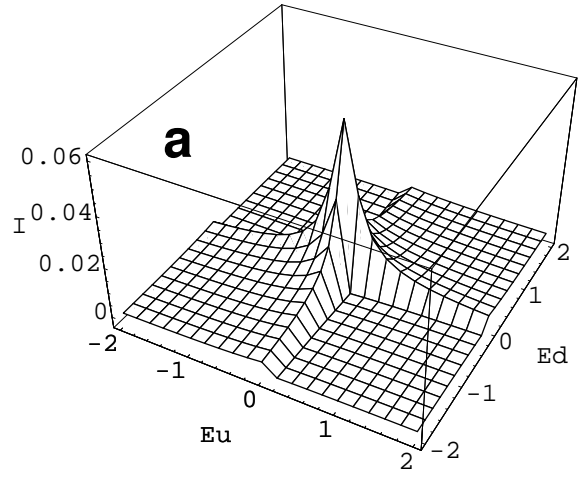
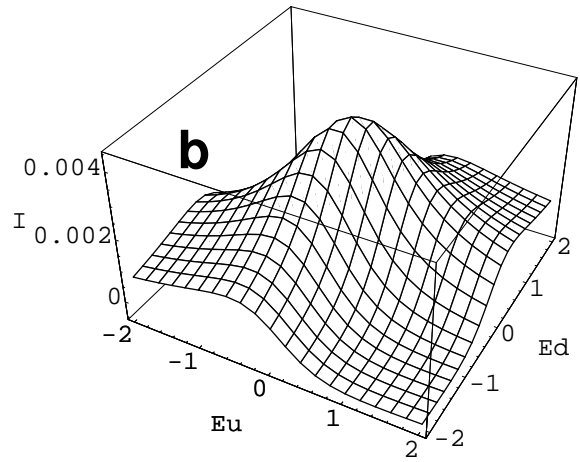
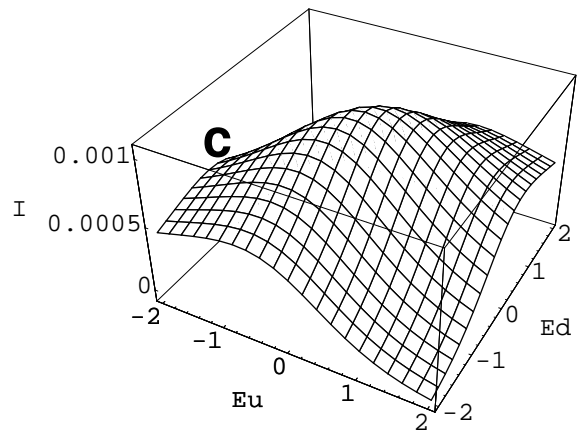
\item[{\bf Fig. 4}]  Surfaces of Josephson current $I$ vs $(E_{\uparrow
},E_{\downarrow })$ configurations at different temperatures: $%
T=10^{-4},\;0.25,\;0.50$ for (a), (b) and (c), respectively. Other
parameters are the same as Fig.1.

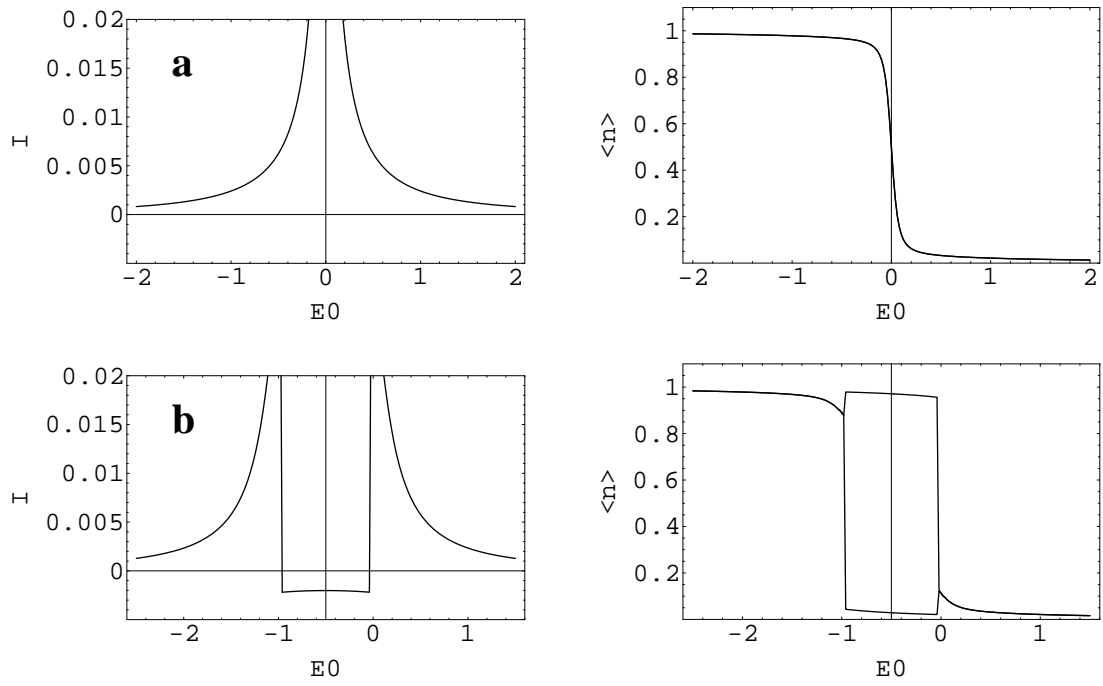
\item[{\bf Fig. 5}]  The Josephson current $I$ and the averaged QD
occupation number $\left\langle n_\sigma \right\rangle $ vs the bare QD
level $E_0$, for (a) QD without interaction and (b) QD with strong
interaction. Parameters are: $\Delta =1,\;\Gamma =0.1,\;\phi =\frac \pi 2%
,\;T=0$, $U=0$ for (a) and $U=1$ for (b).

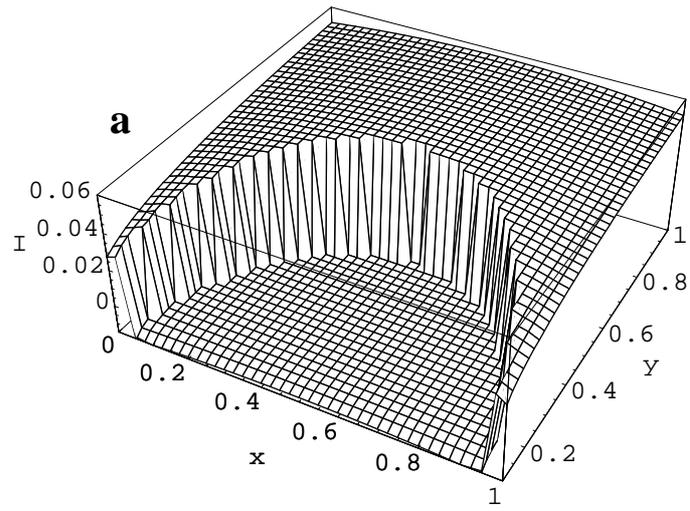
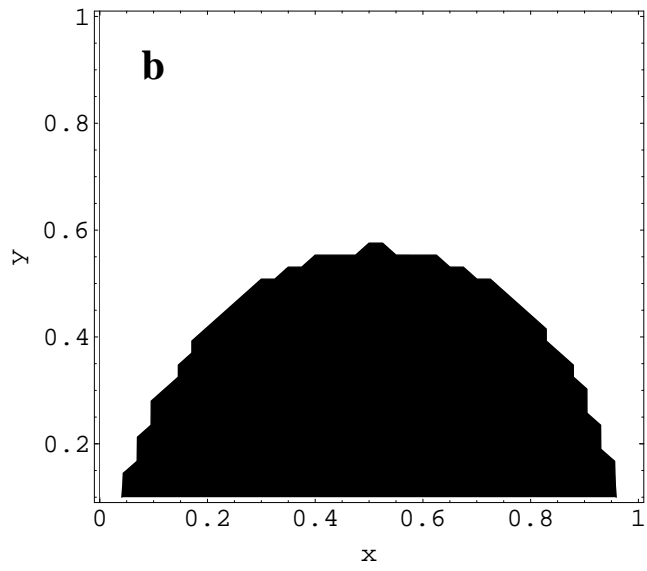
\item[{\bf Fig. 6}]  Parametric diagram of the Josephson current $I$ vs
parameters $x\equiv -E_0/U$ and $y\equiv \Gamma /U$. We set $\Delta
=1,\;\Gamma =0.1,\;\phi =\frac \pi 2,\;T=0$, and change $E_0$ and $U$ to
obtain the surface graph in (a). The black area of (b) indicates the range
of parameter where the S-QD-S system behaves as a $\pi $-junction.

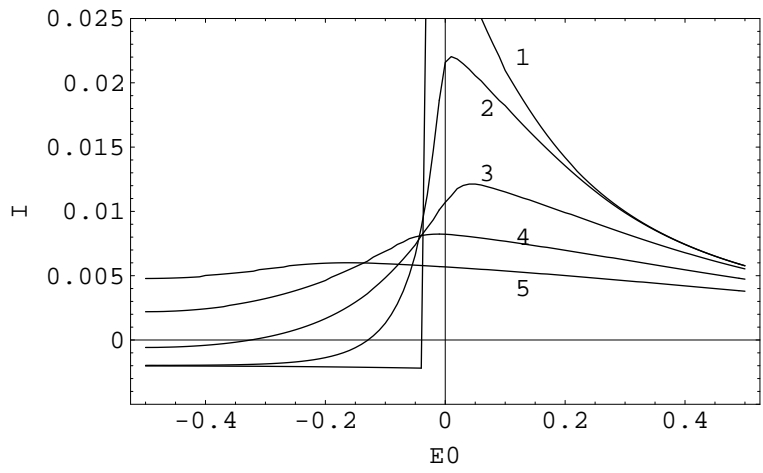
\item[{\bf Fig. 7}]  Temperature dependence of the $I$ vs $E_0$. Parameters
are: $\Delta =1,\;\Gamma =0.1,\;\phi =\frac \pi 2,\;U=1$; $%
T=0.001,\;0.05,\;0.10,\;0.15$, and$\;0.20$ for curves marked by 1, 2, 3, 4,
and 5, respectively.

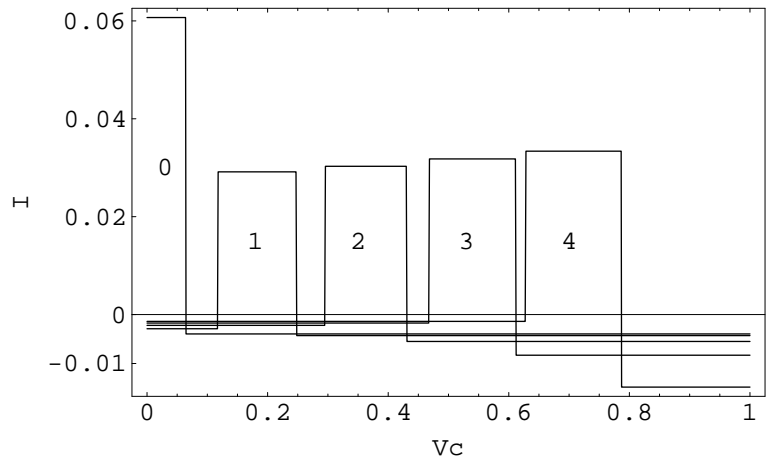
\item[{\bf Fig. 8}]  The Josephson current $I$ vs the control voltage $V_c$
for different Zeeman splitting of $E_{\uparrow }=-E_{\downarrow }=h$ in QD.
Parameters are: $\Delta =1,\;\Gamma =0.1,\;\phi =\frac \pi 2,\;T=0$; $%
h=0,\;0.2,\;0.4,\;0.6$, and $0.8$ for curves marked with 0, 1, 2, 3, and 4,
respectively.
\end{itemize}


\begin{references}
\bibitem{sup}  A. F. Morpurgo, T. M. Klapwijk, and B. J. van Wees, Appl.
Phys. Lett. {\bf 72, }966 (1999).

\bibitem{enh}  J. Kutchinsky, R. Taboryski, C. B. S$\phi $rensen, J. B.
Hansen, and P. E. Lindelof, Phys. Rev. Lett. {\bf 83, }4856 (1999).

\bibitem{rev}  J. J. Baselmans, A. F. Morpurgo, T. M. Klapwijk, and B. j.
van Wees, Nature (London) {\bf 43, }397 (1999).

\bibitem{arm1}  H. Nakano and H. Takayanagi, Phys. Rev. B {\bf 47, }7986
(1993).

\bibitem{arm2}  H. T. Ilhan, H. V. Demir, and P. F. Bagwell, Phys. Rev. B 
{\bf 58, }15120 (1998).

\bibitem{btk}  G. E. Blonder, M. Tinkham, and T. M. Klapwijk, Phys. Rev. B 
{\bf 25, }4515 (1982).

\bibitem{bll1}  B. J. van Wees, K.-M. H. Lenssen, and C. J. P. M. Harmans,
Phys. Rev. B {\bf 44, }470 (1991).

\bibitem{bll2}  G. Wendin and V. S. Shumeiko, Phys. Rev. B {\bf 53, }R6006
(1996).

\bibitem{bll3}  L.-F. Chang and P. F. Bagwell, Phys. Rev. B {\bf 55, }12678
(1997).

\bibitem{abs}  P. F. Bagwell, Phys. Rev. B {\bf 46, }12573 (1992).

\bibitem{dff1}  A. F. Volkov, Phys. Rev. Lett. {\bf 74, }4730 (1995).

\bibitem{dff2}  A. F. Volkov and H. Takayanagi, Phys. Rev. B {\bf 56, }11184
(1997).

\bibitem{dff3}  S.-K. Yip, Phys. Rev. B {\bf 58, }5803 (1998).

\bibitem{dff4}  F. K. Wilhelm, G. Sch\"{o}n, and A. D. Zaikin, Phys. Rev.
Lett. {\bf 81, }1682 (1998).

\bibitem{sds1}  L. I. Glazman and K. A. Matveev, JETP Lett. {\bf 49, }659
(1989).

\bibitem{sds2}  B. I. Spivak and S. A. Kivelson, Phys. Rev. B {\bf 43, }3740
(1991).

\bibitem{sds3}  S. Ishizaka, J. sone, and T. Ando, Phys. Rev. B {\bf 52, }%
8358 (1995).

\bibitem{sds4}  A. V. Rozhkov and D. P. Arovas, Phys. Rev. Lett. {\bf 82, }%
2788 (1999).

\bibitem{sds5}  A. A. Clerk and V. Ambegaokar, cond-mat/9910201.

\bibitem{sfs1}  V. Proki\'{c}, A. I. Buzdin, and L.
Dobrosavljevi\'{c}-Gruji\'{c}, Phys. Rev. B {\bf 59, }587 (1999).

\bibitem{sfs2}  S.-K. Yip, Phys. Rev. B {\bf 62, }R6127 (2000).

\bibitem{sfs3}  T. T. Heikkil\"{a}, F. K. Wilhelm, and G. Sch\"{o}n,
cond-mat/0003383.

\bibitem{ar3}  Q.-F. Sun, B.-G. Wang, and T.-H. Lin, Phys. Rev. B {\bf 61, }%
4754 (2000).

\bibitem{ar4}  Q.-F. Sun, J. wang, and T.-H. Lin, Phys. Rev. B {\bf 62, }648
(2000).

\bibitem{nbrp}  J. C. Cuevas, Mart\'{i}n-Rodero, and A. L. Yeyati, Phys.
Rev. Lett. {\bf 29, }3486 (1997).

\bibitem{mag}  P. W. Anderson, Phys. Rev. {\bf 124, }41 (1961).

\bibitem{add1}  A. D. Zaikin and G. F. Zharkov, Sov. Phys. JETP {\bf 51, }%
364 (1980).

\bibitem{add2}  A. V. Tartakovskii and M. V. Fistul, Sov. Phys. JETP {\bf %
67, }1935(1988).

\bibitem{remark1}  The phase factor is $e^{-\text{i}\phi }$ instead of $e^{%
\text{i}\phi }$ since we choose $e=1$ instead of $e=-1$.

\bibitem{remark2}  The contribution from the discrete spectrum can also be
derived from the dipersion relation of Andreev bound states, $I_d=2\left[ 
\frac{\partial \tilde{E}_1}{\partial \phi }f(\tilde{E}_1)+\frac{\partial 
\tilde{E}_2}{\partial \phi }f(\tilde{E}_2)\right] $.
\end{references}
\end{document}